\documentstyle[prl,aps,epsfig]{revtex}

\begin{document}
\large

\title{Quantum information processing by NMR using a 5-qubit system formed by dipolar coupled spins in an oriented molecule}
\author{Ranabir Das$^\dagger$, Rangeet Bhattacharyya$^\dagger$ and 
Anil Kumar $^{\dagger \ddagger}$\footnote{ {\small \it DAE-BRNS Senior Scientist}\\}\\
        $^{\dagger}$ {\small \it Department of Physics, Indian Institute of Science, Bangalore, India}\\
        $^{\ddagger}$ {\small \it Sophisticated Instruments Facility, Indian Institute of Science, Bangalore, India}\\}

\maketitle
\vspace{0.5cm}
\begin{abstract}
Quantum Information processing by NMR with small number of qubits is well established. Scaling to
higher number of qubits is hindered by two major requirements (i) mutual coupling among qubits and
(ii) qubit addressability. It has been demonstrated that mutual coupling can be increased by using
residual dipolar couplings among spins by orienting the spin system in a liquid crystalline matrix.
In such a case, the heteronuclear spins are weakly coupled but the homonuclear spins
become strongly coupled. In such circumstances, the strongly coupled spins can no longer be treated as 
qubits. However, it has been demonstrated elsewhere, that the $2^N$ energy levels of a strongly coupled N spin-1/2 system can  
be treated as an N-qubit system. For this purpose the various transitions have to be 
identified to well defined energy levels. This paper consists of two parts. In the first part, the 
energy level diagram of a heteronuclear 5-spin system is obtained by using a 
 newly developed heteronuclear z-cosy (HET-Z-COSY) experiment.
In the second part,  implementation of logic gates, preparation of pseudopure states, creation of entanglement and 
entanglement transfer is demonstrated, validating the use of such systems for quantum information processing.
\end{abstract}

\section{Introduction}

 Theoretically, future quantum computers  can simulate physical systems and 
solve certain problems  more efficiently than classical computers \cite{rf,deu,grover,shor,preskill}.
This possibility has excited several researchers to strive for experimental realization of
 quantum computers. Apart from trapped ions and cavity-QED systems, Nuclear Magnetic Resonance (NMR) 
has proved as a suitable physical system for quantum information processing (QIP) 
\cite{chuangbook,bouwmeester}. In fact, NMR has emerged as the 
most successful of all systems, demonstrating several quantum algorithms 
including Shor's factorization algorithm \cite{djchu,djjo,free,grochu,kavita00,nat,ranapra}. 
However, scalability is one of the hurdles that NMR-QIP faces today.   

Originally it was conceived that liquid-state NMR will use as qubits, spin-1/2 nuclei having different 
Larmor frequencies and weakly coupled to each other by indirect J-couplings \cite{cory97,chuang97,cory98}. 
Since J-couplings mediate through covalent bonds, they become vanishingly small beyond 4-5 bonds. 
Such systems are thus difficult to scale beyond 7-8 qubits. 
Attempts are going on to explore other couplings such as the direct dipolar couplings between spin-1/2 nuclei,
 which are larger in magnitude, 
depend on the inter-nuclear distances, and not on the number of intermediate covalent bonds. 
However, in liquids, molecules undergo fast isotropic reorientations, 
thereby averaging out all dipolar interactions ( homonuclear, heteronuclear, intermolecular and intra molecular). 
In solids, the molecules are static, and all the dipolar interactions are retained 
yielding broad spectra, rendering such systems unsuitable for quantum information processing.
However, when molecules are dissolved in liquid crystal solvents, they have sufficient translational motion but get partially oriented
 and have anisotropic rotational motion.
In such cases, while the  intermolecular dipolar interactions are averaged to zero, the
 intra-molecular dipolar interactions survive which are scaled down from 
several KHz to a few KHz.           
Such systems yield  well resolved spectra with finite number of sharp lines \cite{khetrapal}. 
In such cases the spins have considerable 
dipolar couplings and it has been demonstrated that they can be used for quantum information 
processing \cite{chuang00b,marka,fung00,mahesh02}.

However, in cases where the differences in Larmor frequencies or chemical shifts
are not large compared to the dipolar couplings, the spins become strongly coupled. The eigenstates of the 
system are not products of the eigen-states of various spins but linear combinations of the product states, and 
spins can no longer be identified as a qubits \cite{ernstbook}. Still, it has been demonstrated that
 the $2^N$ energy levels of the N spin-1/2 system can be treated  
as $2^N$ states of a N-qubit system, and, can be used for quantum information processing \cite{strong}. 
Quantum Information processing has already been demonstrated in homonuclear 2,3 and 4-spin 
 systems oriented in liquid crystal matrices \cite{strong}. In the homonuclear systems, it was found that   
if the  spins are in symmetric chemical environment, 
the eigenstates break up into symmetry groups, leading to reduced number of qubits 
and other complications with degenerate transitions from different symmetry manifolds \cite{fung00,mahesh02}.
It is therefore desirable that the homonuclear spins are in non-symmetric chemical environment.
 Furthermore, it is advantageous to include heteronuclear spins in the system. 
Here, we extend our earlier work by including a heteronuclear spin with 
four homonuclear spins in non-equivalent chemical environment, all dipolar coupled to each other. We use 
the molecule   1-chloro,2-flouro benzene 
(dissolved in liquid crystal solvent ZLI-1132), which yields a 5-qubit system in which the four protons 
are strongly coupled to each other but weakly coupled to the fluorine. The Hamiltonian for the 
dipolar interaction among the strongly coupled protons $(\bf{I_i})$ is,
\begin{eqnarray}
H_D= 2\pi \sum_{i,j(i<j)} D_{ij}^{ori} (3I_{iz}I_{jz}-{\bf I_i.I_j}),
\end {eqnarray}
where
$D_{ij}^{ori}$=$\frac{\gamma_i\gamma_j \hbar}{4\pi r^3_{ij}}  \Omega_{ij}$ is the scaled dipolar interaction, 
scaled by the order parameter $\Omega_{ij}$ \cite{khetrapal}. The 
Hamiltonian for dipolar interaction between the protons and the fluorine({\bf S}) is 

\begin{eqnarray}
H_D= 2\pi \sum_{i} D_{I_iS}^{ori} I_{zi}S_z,
\end {eqnarray}
The full Hamiltonian of the system including Zeeman, dipolar and J-coupling terms, is given by:
\begin{eqnarray}
H&=&H_Z+H_D+H_J= \sum_{i}\omega_{I_i}I_{zi} + \omega_sS_{z} \nonumber \\
&&+2\pi\sum_{i,j(i<j)} D_{ij}^{ori} (3I_{iz}I_{jz}-{\bf I_i.I_j}) + 2\pi\sum_{i} D_{I_iS}^{ori} I_{zi}S_z 
  +2\pi\sum_{i,j(i<j)} J_{ij}{\bf I_i.I_j}+2\pi \sum_{i}J_{I_iS}I_{zi}S_z.
\end {eqnarray}

 The equilibrium spectrum of the protons 
and the fluorine is given in Fig. 1. 
In order to use this spin system for QIP, one needs to identify the energy levels of the system and to 
assign each transition to a pair of these levels. There are two possible methods. One well-known method 
 is to numerically diagonalize the Hamiltonian of the system, with some initially guessed parameters 
(chemical shifts, dipolar and J-couplings) and iteratively fit the calculated and observed spectra \cite{khetrapal}. 
The second method is the use of Z-COSY two-dimensional (2D) experiment which identifies connected transitions 
and draws automatically the energy-level diagram \cite{zcosy1}. 
So far Z-COSY has been used only for homonuclear spins. 
This paper consists of two parts. In the first part (section II) we introduce a heteronuclear Z-COSY experiment 
and utilize it to obtain the complete energy level diagram of this dipolar coupled 5 spin system. 
In the second part (section III), we demonstrate that this spin system can be used as a 5-qubit 
system by preparing pseudopure states, performing C$^n$-NOT operations and controlled SWAP operations, 
and creating and transferring entanglement using transition selective pulses.

\section{HET-Z-COSY} 
The pulse sequence
of a Z-COSY experiment is $90-t_1-\alpha-\tau-\beta-t_2$, where only the longitudinal magnetization is
retained during the interval $\tau$ by either phase cycling or by a 
gradient pulse \cite{zcosy1,zcosy2}. The 90$^o$ pulse converts the
equilibrium z-magnetization into coherences, which are frequency labeled during the period $t_1$.
 The small angle $\alpha$ pulse ensures that each cross-section parallel to $\omega_2$ from the resulting 2D spectrum is
equivalent to a one-dimensional (1D) experiment in which the peak corresponding to the diagonal is selectively inverted.
The directly connected transitions to the inverted transition are finally measured in the linear regime 
by a small angle $\beta$ pulse. The
Z-COSY experiment has been successfully used to construct the energy level diagram in oriented homonuclear spin 
systems \cite{zcosy1,surya}.
Here the Z-COSY experiment is extended to include a heteronucleus  and we call this experiment as HET-Z-COSY.

For ``n" heteronuclear systems HET-Z-COSY experiment requires ``n" 2D experiments where each experiment
 observes one of the heteronucleus during $t_2$. The pulse sequence for HET-Z-COSY experiment is 
$90-t_1-\alpha-(Gz+kt_1)-\beta-t_2$, where the $90^o$ and $\alpha$ pulses are applied on all the nuclei, and 
$\beta$ is applied only on the observed nucleus.  
 In the present case, two experiments have to be performed for detecting protons and fluorine respectively.
 Gz is the gradient pulse which retains only the 
the longitudinal magnetization. Besides creating multi-spin zz terms \cite{zcosy1}, the $\alpha$-pulse creates all 
coherences including zero-quantum coherences. The gradient pulse destroys all other coherences as desired, 
except the homonuclear zero-quantum coherences. These zero-quantum coherences get 
partially converted into single quantum coherences by the last $\beta$ pulse causing the well known zero-quantum 
interference in the 2D spectroscopy. To avoid this we have used the method of zero-quantum shifting \cite{zcosy2} by 
incrementing the $\tau$ delay as (Gz+kt$_1$), where k=2 was used. This shifts the zero-quantum peaks in the 
$\omega_1$ dimension which are eventually removed by symmetrizing the spectrum \cite{zcosy1}. 

The 2D spectrum of HET-Z-COSY on the systems of oriented 
1-chloro 2-flouro benzene is shown in Fig. 2. The experimental spectrum when protons are observed in 
$t_2$, is given on the left side ((a) and (c)), while the experimental spectrum when fluorine is observed
 is given on the right  side ((b) and (d)). The spectrum in figure 2(a) has the diagonal and connectivity 
information between the proton transitions, while figure 2(c) has the connectivity
information of the fluorine transitions to the proton transitions. Figure 2(b)   
 has the diagonal and connectivity information between the fluorine transitions, and figure 2(d) 
has the connectivity information of the proton transitions to the fluorine transitions. 
Every connectivity is mutual such that if transition $`a'$ is progressively connected to transition $`b'$, 
then transition $`b'$ is also progressively connected to transition $`a'$. 
Hence the cross peaks should be symmetric about the diagonal \cite{zcosy1}. 
The symmetrized spectrum (free of zero-quantum and other artifacts) is given in Fig. 3. A connectivity matrix is 
constructed from the symmetric 2D spectrum, by taking cross-sections parallel to $\omega_2$ at each transition along 
$\omega_1$. From this connectivity matrix the complete energy-level diagram can be constructed following the 
method given in reference 32 \cite{zcosy1,surya}. 
The complete energy level 
diagram constructed by the above procedure is given in fig. 4. All the steps including construction of 
the juxtaposed 2D spectrum, symmetrization, construction of the connectivity matrix and the construction of 
the complete energy-level diagram are 
automated by a program written in Matlab \cite{prog}.
   
In this 5-spin system, there are $2^5=32$ energy levels. The proton transitions connect energy levels 
within two separate domains (A and B) of 16 energy levels each, corresponding respectively to +1/2 and -1/2 states 
of the fluorine spin (figure 4). The fluorine transitions connect the levels from A domain to the B domain. 
 The various transitions 
observed in the spectrum are given serial numbers (1-60 for proton and 61-82 for the fluorine transitions). 
Except the four extreme eigenstates all others are linear combinations of the product states of the 4 protons. 
The extreme eigenstates are 1A, 16A, 1B and 16B respectively corresponding to the eigenstates
 $\vert \alpha\alpha\alpha\alpha\alpha\rangle$, 
$\vert \alpha\beta\beta\beta\beta\rangle$, $\vert \beta\alpha\alpha\alpha\alpha\rangle$ and 
$\vert \beta\beta\beta\beta\beta\rangle$, where the first state corresponds to the fluorine spin and last 
four to the protons.  

\section{Quantum Information Processing using transition selective pulses}
 Fig. 5 shows the labeling of the various eigenstates of the 5-spin system to 5-qubit labels. While labels 
can be chosen for optimum execution of various logical operations, 
here we have retained a conventional 
labeling \cite{ranapra1}.  Transition selective $(\pi)$-pulse has a narrow bandwidth 
and when tuned to the resonance frequency of a 
specific transition in the spectrum, it  exchanges the amplitudes of 
the two eigen-states which are connected by that transition \cite{ernstbook}. Such pulses have already been used to simplify the
implementation of several logical operations for quantum information processing 
\cite{free,kavita00,cory98,fung00,maheshpra01,peng,lo,ranajmr}.

\subsection{Preparation of pseudopure states}
Pseudopure states (PPS) are the staring point for implementation of various quantum algorithms by NMR.
In a PPS, populations of all the states except one are equalized \cite{chuang97,cory98}. 
Among the various methods of creating PPS, we have implemented the method of
Pair of pseudopure states (POPS), originally suggested by Fung et.al \cite{fung00}.
POPS requires two experiments:
 (i) equilibrium  and (ii) a selective population inversion. Subtraction of these two 
population distributions yields zero populations for all the levels except the inverted pair, which in turn 
has equal positive/negative populations. This acts as a pair of pseudopure states. It has been demonstrated 
elsewhere, that for implementation of certain algorithms, POPS is sufficient to begin with \cite{fungjcp}.
Moreover, in an N-qubit system, POPS can also act as a (N-1)-qubit sub-system PPS. 

 Here we have chosen to apply 
$(\pi)$ pulse on the fluorine transition 61 which connect  
$\vert 00000 \rangle \leftrightarrow \vert 10000 \rangle$.
 After the $(\pi)^{61}$ pulse, a gradient pulse is applied to kill any unwanted coherence created by the 
imperfection of the r.f. pulse. The final 
populations are measured by a small angle ($\pi/20$) proton and fluorine pulses, which convert within linear approximation,
 population differences into observable single quantum intensities. The resulting spectrum is given in 
figure 6(b). Subtraction of equilibrium populations (figure 6(a)) from that corresponding to 
figure 6(b) yields the  pair of pseudopure states 
$\vert 10000 \rangle \langle 10000 \vert - \vert 00000 \rangle \langle 00000 \vert$. The resulting spectrum 
is given in figure 6(c). 
Note that the transitions 2, 18, 21 and 37 are of opposite sign to that of 61, as they are
progressively connected to 61 (see figure 5) and transitions  1 and 9 are of the same sign since that they are regressively
connected to 61. The spectrum obtained after POPS contains only transitions originating from the two levels having non-zero 
populations. This effectively gives the various transitions connected to these levels. The same information is also available 
in the cross-sections of HET-Z-COSY spectrum at the frequency of transition connecting the two levels. 
 A cross-section parallel to
$\omega_2$ at the $\omega_1$-frequency of transition 61 from spectrum of 
figure 2  is given in figure 6(c'). A good match of relative intensities 
of various transitions within each spectrum of 6(c) and 6(c') confirms the creation of POPS in 6(c).      
It may be noted that, while 6(c) is achieved with a selective inversion of transition 61, 
6(c') is a cross-section of a 2D spectrum obtained using non-selective low angle pulses assuming  perturbation and 
observation in the linear regime.

Another method of creating sub-system pseudopure state is; Spatially Averaged Logical Labeling 
Technique (SALLT) \cite{maheshpra01}. A hard $(\pi/2)$ pulse on the protons equalize the populations within each  
subsystem [A and B], with the value differing between the two subsystems. 
A gradient pulses is subsequently applied to kill the coherences created in the process. 
A transition selective pulse $(\pi)$ on one of the fluorine transitions (we have chosen transition 61) then creates  
subsystem pseudopure states.       
The populations corresponding to this state are measured by a small angle ($\pi/20$) pulse and the resulting spectrum, 
 is given in figure 6(d). The proton spectra of figure 6(c)
and 6(d) are nearly identical, confirming the creation of subsystem PPS. The populations of POPS and SALLT 
differ in the sense that POPS yields  zero populations 
 of all levels except one in each subsystem, 
while SALLT has equal population of all levels (not necessarily zero) except one in each subsystem. The fluorine spectrum 
in figure 6(d) is therefore identical to the spectrum in figure 6(b).  

\subsection{Implementation of controlled-NOT and controlled-SWAP gates}
Controlled-NOT gates are essential parts of any computation. A C$^4$-NOT gate will change the state of
fifth qubit when the other four qubits are in the state $\vert 1 \rangle$. Proton transition 60 connects the states
$\vert 11110 \rangle$ and $\vert 11111 \rangle$ (see fig. 5). Hence a $(\pi)^{60}$ pulse
performs C$^4$-NOT operation. We start from equilibrium state and implement the C$^4$-NOT gate. A subsequent gradient
pulse kills any coherences created by imperfection of the r.f pulse. The output
of the gate is stored in the final populations. The final populations, measured by  small-angle $(\pi/20)$ proton and fluorine pulses
 after implementing C$^4$-NOT gate corresponding to
$\vert 11111\rangle \leftrightarrow \vert 11110\rangle$, yield the spectrum given in figure 7(b).
 Another POPS can be created by subtraction of this population distribution and the equilibrium populations. 
Subtraction of figure 7(a) from 7(b) yields the POPS of $\vert 11111 \rangle \langle 11111 \vert - \vert 11110 \rangle \langle 11110 \vert$.
The corresponding spectrum is  
given in figure 7(c). Creation of the POPS can also be confirmed by cross-sections parallel to 
$\omega_2$ of figure 2 at the $\omega_1$-frequency of the transition 60, given in 7(c').

Controlled SWAP gates are becoming increasingly popular in quantum information processing \cite{gate,gerry}. 
A C$^3$-SWAP gate swaps the state of last two qubits when the first three qubits are in the state 
$\vert 1 \rangle$. Essentially, this gate interchanges the
amplitudes between $\vert 11110 \rangle$ and $\vert 11101 \rangle$. A sequence of three 
pulses $(\pi)^{58}$-$(\pi)^{52}$-$(\pi)^{58}$ achieves this gate. 
We start from the POPS of 
$\vert 11111 \rangle \langle 11111 \vert - \vert 11101 \rangle \langle 11101 \vert$ prepared by 
subtraction of the equilibrium populations from the populations obtained by inversion of the proton transition 55 (figure 7(d)).
The creation of this POPS is confirmed by the cross-section of figure 2 at the frequency of 55, given in figure (d').
 The subsequent implementation of C$^3$-SWAP gate [$(\pi)^{58}$-$(\pi)^{52}$-$(\pi)^{58}$] interchanges the 
populations between $\vert 11110 \rangle$ and $\vert 11101 \rangle$. A gradient pulse is applied after each $(\pi)$-pulse to 
destroy any coherences created by imperfection of pulses. The final state is the 
POPS of $\vert 11111 \rangle \langle 11111 \vert - \vert 11110 \rangle \langle 11110 \vert$, and the resulting proton and 
fluorinne spectra measured by a small angle ($\pi/20$) pulses is given in figure 7(e). 
A good match between 7(e) and 7(c) confirms the implementation of 
C$^3$-SWAP gate.     

\subsection{Entanglement creation and transfer}
Entanglement is a remarkable property of quantum systems, which has no classical analogue. 
Entanglement is also a key requirement for many quantum information processing protocols. 
Many of these protocols require transfer of entanglement between qubits. Entanglement transfer has 
become specially useful in the cases where one needs to transfer data to the  quantum memory 
(qubits with less decoherence that are used for storing data) 
from the fast processor (qubits have decoherence but can perform  universal logic operations) in a manner similar 
to the protocol 
proposed by Chuang et.al. \cite{got}. Entanglement transfer has been demonstrated previously by Cory et.al. in 
liquid state NMR of weakly coupled 4-qubit spin-1/2 system \cite{enttr}. Here, we demonstrate entanglement transfer 
in this strongly coupled system by using the last 4-qubits of the 5-qubit system with the first qubit acting as an  
ancillary bit. We start by preparing the 4-qubit subsystem pseudopure state by SALLT (figure 6(d)). 
We chose to perform further operations in only one of the subsystems (where 1st qubit is in state $\vert 1\rangle$)
 [domain B]. Hence, in the next paragraph, only the states of the last four qubits are given, but they are to be understood 
as the states with first qubit being $\vert 1\rangle$. 

 Starting from the $\vert 0000 \rangle$ pseudopure state, uniform superposition of 2nd qubit was created 
by a $(\pi/2)^{2}$ pulse: $\vert 0000 \rangle \rightarrow \vert 0000 \rangle+\vert 0100 \rangle$ (fig 5(a), B domain).    
The states were tomographed using two-dimensional Fourier transform technique \cite{ranabirtomo}.  
2D spectra for measurement of all off-diagonal elements of the density matrix are given in figure 8. 
Figure 8(a) contains the totality of the multiple quantum (MQ) spectrum of the four protons obtained 
by a $(\pi/2)-\tau-(\pi/2)$ pulse sequence starting from equilibrium population distribution. 
Figure 8(b) on the other hand contains the MQ spectrum starting from the 
state of $\vert 0000 \rangle+\vert 0100 \rangle$ and shows 
the presence of a single quantum coherence 
of this state at frequency of transition '2'( $\omega^{(2)}$) along $\omega_1$.
 A controlled not gate between 2nd and 3rd qubit 
is implemented by a $(\pi)^{27}$ pulse. This gate changes the state of third qubit when the second qubit 
is in state $\vert 1 \rangle$ and  creates an entangled 
EPR (Einstein-Podolsky-Rosen \cite{epr}) pair between the 2nd and 3rd qubit given by:
$(\vert 0000 \rangle+\vert 0110 \rangle)$. 
The presence of the double quantum coherence with frequency $\omega^{(2)}+\omega^{(27)}$ along $\omega_1$
in the 2D spectrum for tomography of all off-diagonal elements in figure 8(c) 
and the absence of all other coherences, indicates the creation of the EPR coherence.
The tomographed density matrix of the B domain, given in figure 8(d), confirm creation of the EPR state.     
A sequence of three transition selective pulses $(\pi)^{27}-(\pi)^{39}-(\pi)^{27}$ swaps the states
 of $\vert 0110 \rangle$ and $\vert 1001 \rangle$, thereby transferring the entanglement
to the first and fourth qubit: $(\vert 0000 \rangle+\vert 1001 \rangle)$. 
Hence the entanglement initially created between the second and third qubit is now transfered to the 
first and fourth qubit, while the second and third qubit are in the  
$\vert 00 \rangle$ state and can be used for further computation. 
All the pulses and the receiver were phase cycled by a standard CYCLOPS phase cycle \cite{ernstbook}.
The presence of the double quantum coherence with frequency $\omega^{(2)}+\omega^{(39)}$ along $\omega_1$
in the 2D spectrum in figure 8(e) and the resulting tomograph of the B domain, given in figure 8(f), 
confirm that the entanglement has been transferred with 85$\%$ fidelity.
\section{conclusion}
Residual dipolar couplings can be used to increase the number of qubits for quantum information processing. 
We have used a molecule containing 5 spins oriented in a liquid crystal matrix, and exploited the residual dipolar coupling 
 to demonstrate it as a 5-qubit quantum computer. Transition assignment is performed using HET-Z-COSY experiment 
and qubit addressability is achieved by transition selective pulses. It is expected that using this protocol
higher qubits can be achieved. Preliminary experiments promise a 8-qubit system \cite{nmrs2004}.

 For the implementations reported 
in this paper, evolution under the internal Hamiltonian was not explored. It is however interesting 
to investigate how effectively the evolution under internal Hamiltonian can be manipulated 
to implement quantum algorithms in these systems. Efforts are ongoing in this direction in our laboratory, 
and recently a new method of preparing pseudopure states in oriented systems by 
exciting selected multiple quantum using evolution under 
effective dipolar Hamiltonian, has been reported \cite{khit}.  
\section{Acknowledgments}
 The authors thank T.S. Mahesh, N. Suryaprakash and K.V. Ramanathan
for useful discussions. The use of DRX-500 NMR spectrometer funded by the Department of
Science and Technology, New Delhi, at the Sophisticated
Instrument's Facility, Indian Institute of Science, Bangalore, is also gratefully acknowledged.
AK acknowledges "DAE-BRNS" for the award of "Senior Scientists scheme",
and DST for a research grant on "Quantum Computing using NMR techniques".


\pagebreak
Figure Captions

Figure 1: Equilibrium spectrum of 1-chloro,2-flouro benzene
dissolved in liquid crystal solvent ZLI-1132. (a) The spectrum of four strongly coupled protons . (b) The 
fluorine spectrum. There are 82  
 transitions with significant intensities of which 1-60 are proton transitions and 61-82 
are fluorine transitions. One fluorine transition did not show any connectivity in the subsequent HET-Z-COSY spectra 
and could not be assigned to any energy level in the energy-level diagram. This transition is shown by asterisk (*).\\

Figure 2: Phase-sensitive HET-Z-COSY spectrum of oriented 1-chloro,2-flouro benzene. (a) and (c) is  obtained  
when protons are observed in $t_2$ while  
(b) and (d) is obtained when fluorine is observed in $t_2$. 2K$\times$2K data points were collected with 
8 scans per $t_1$ increment. The data was zero-filled to 4K$\times$4K before Fourier transform.
 The spectrum in (a) has the diagonal and connectivity
information between proton transitions, while that in (c) has the connectivity
information of the fluorine transitions to proton transitions. (b)
 has the diagonal and connectivity information between fluorine transitions, and (d)
has the connectivity information of the proton transitions to fluorine transitions. Artifacts 
 due to incomplete suppression of axial and double quantum diagonal are also visible in the spectrum 
. These artifacts are removed on symmetrization (fig. 3) \\

Figure 3: Symmetric 2D spectrum of  figure 2. 
A peak picking was performed on the 2D spectra of figure 2. The data from the two experiments 
were added to create a single 2D spectrum. This 2D spectrum was then symmetrized by a program written in 
Matlab \cite{prog}. The symmetrization is performed by retaining the lower absolute value 
of ($\omega_i, \omega_j$) and ($\omega_j, \omega_i$) cross-peaks, at both the cross-peak positions. 
For better visibility, the symmetric 2D spectrum is plotted with identical absolute intensity for all the peaks.
The positive peaks are shown by filled circles  
while the negative peaks are shown by empty circles.\\
 
Figure 4: Energy-level diagram of the 5 spin system of oriented 1-chloro,2-flouro benzene. There are 
32 ($2^5$) energy levels which can be divided 
into two separate domains, 
A and B of 16 energy levels each. Domain A contains energy levels with fluorine in +1/2 state and  
domain B contains energy levels with fluorine in -1/2 state.
Proton transitions connect energy levels within the two domains      
whereas fluorine transitions connect energy levels between A and B.\\

Figure 5: Labeling of the eigenstates of 5-spin system to 5-qubit labels. The first qubit is in state 
$\vert 0\rangle$ in domain A, whereas it is in state $\vert 1 \rangle$ in domain B. We have followed 
a conventional labeling scheme here such that in the limit of weak coupling, the state $\alpha (+1/2)$ is given label 
'0' and the state $\beta (-1/2)$ is given the label '1'.\\  

Figure 6: Preparation of pseudopure states in the 5-qubit system of 
 oriented 1-chloro,2-flouro benzene. 
 All spectra were obtained by using a  readout small-angle $(\pi/20)$ proton or fluorine pulse. (a) Equilibrium 
$^1$H and $^{19}$F spectra.   
(b) Spectra corresponding to inversion of population between the states $\vert 00000 \rangle$ and $\vert 10000 \rangle$ 
implemented by a selective $(\pi)$ on transition 61. 
(c) Spectra obtained by subtraction of (a) from (b) yielding POPS(61) of 
$\vert 10000 \rangle \langle 10000\vert-\vert 00000 \rangle \langle 00000\vert$  
. Note that the transitions 2, 18, 21 and 37 are of opposite sign to that of 61, confirming that they are 
progressively connected to 61 (see figure 5). 
The transitions  1 and 9 are of the same sign implying that they are regressively 
connected to 61. Transition 19 is of very small intensity and not visible in (d).  
(c') Cross-section parallel to $\omega_2$ of the spectrum in figure 2 at the $\omega_1$-frequency of transition 61.
(d) Spectra corresponding to the 4-qubit subsystem pseudopure state 
by SALLT (see text for details).\\ 

Figure 7: Implementation of logic operations in 5-qubit system of
 oriented 1-chloro,2-flouro benzene.
(a) Equilibrium $^1$H and $^{19}$F spectra.
(b) Spectra corresponding to C$^4$-NOT gate, implemented by a selective $(\pi)$ on transition 60, which
connects $\vert 11110\rangle \leftrightarrow \vert 11111\rangle$ (see figure 5).
(c) Spectra corresponding to the creation of POPS(60) $\vert 11111 \rangle \langle 11111\vert-\vert 11110 \rangle \langle 11110 \vert$
 by subtraction of (a) from (b). In these spectra, transitions 14, 38, 52 and 80 are of opposite 
sign to that of 60, signifying that they are
progressively connected to 60, while 55 and 82 are of the same sign confirming that they are regressively connected.
The transition 45 is of very low intensity and not visible in (c) and (c').

(c') Cross-section parallel to $\omega_2$ of figure 2 at the $\omega_1$-frequency of transition 60.
(d) POPS(55) of $\vert 11111 \rangle \langle 11111\vert-\vert 11101 \rangle 11101\langle \vert$ created by 
a selective $(\pi)$ pulse on 55, and a subsequent subtraction of equilibrium. 
Transitions  33, 58,67, 73 and 75 are of opposite sign to that of 55, signifying that they are
progressively connected to 55 (see figure 5).
The transitions 60 and 82 are of the same sign confirming that they are regressively
connected to 55. Transition 45 is of very low intensity and not visible in (f).
(d') Horizontal cross-section of figure 2 at the frequency of transition 55.
(e) Implementation of C$^3$-SWAP gate on the POPS of (a). A sequence of three selective 
pulses $(\pi)^{58}$-$(\pi)^{52}$-$(\pi)^{58}$ implements the C$^3$-SWAP gate thereby creating the  
 POPS of $\vert 11111 \rangle \langle 11111\vert-\vert 11110 \rangle \langle 11110\vert$.
 A good match between (e) and (c) confirms implementation of C$^3$-SWAP gate.\\

Figure 8: Entanglement transfer within the  last 4-qubits of the 5-qubit system of
 oriented 1-chloro,2-flouro benzene. In all 2D experiments a date matrix of 1K$\times$1K
were collected all zero-filled to 2K$\times$2K before Fourier transform.
 (a) Starting from equlibrium, multiple quantum spectrum of the four protons obtained by a pulse sequence of
$(\pi/2)-\tau-(\pi/2)$ where $\tau=30ms$, followed by the pulse sequence of tomography \cite{ranabirtomo}. 
(b) Spectrum corresponding to measurement of all off-diagonal elements of the state 
$\vert 0000\rangle +\vert 0100\rangle$. A single coherence along $\omega_1 $ at 
the frequency of transition 2,( $\omega^{(2)}$)  and absence of any other coherences confirm  
creation of this state.  (c) Spectrum corresponding to measurement of all off-diagonal elements of the entangled state
$\vert 0000\rangle +\vert 0110\rangle$. A double quantum coherence along $\omega_1 $ at
the frequency  $\omega^{(2)}+\omega^{(27)}$  and absence of any other coherences confirm
creation of EPR coherences. (d) Corresponding tomographed density matrix using the method 
outlined in \cite{ranabirtomo}.
(e) Spectrum corresponding to measurement of all off-diagonal elements of the state
after entanglement transfer. A double-quantum coherence is observed along $\omega_1 $ at
the frequency  $\omega^{(2)}+\omega^{(40)}$ with a small coherence at $\omega^{(2)}+\omega^{(27)}$.  
(f) Corresponding tomographed density matrix confirm transfer of entanglement with 85$\%$ fidelity.   
   
\pagebreak
\begin{figure}
\epsfig{file=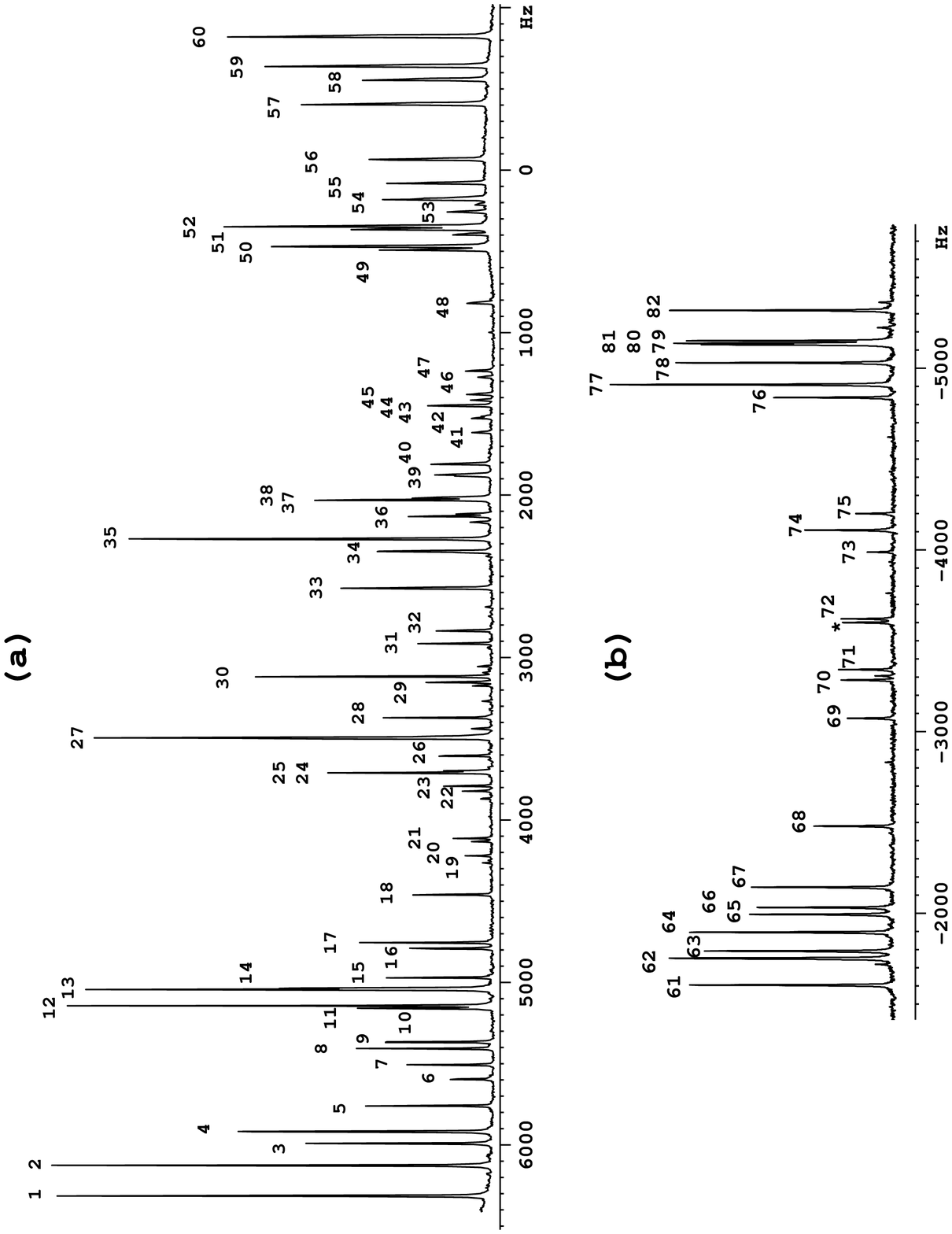,height=20cm}
\end{figure}
\hspace{5cm}
{\huge Figure 1}

\pagebreak
\vspace*{-20cm}
\begin{figure}
\hspace{-1.5cm}
\epsfig{file=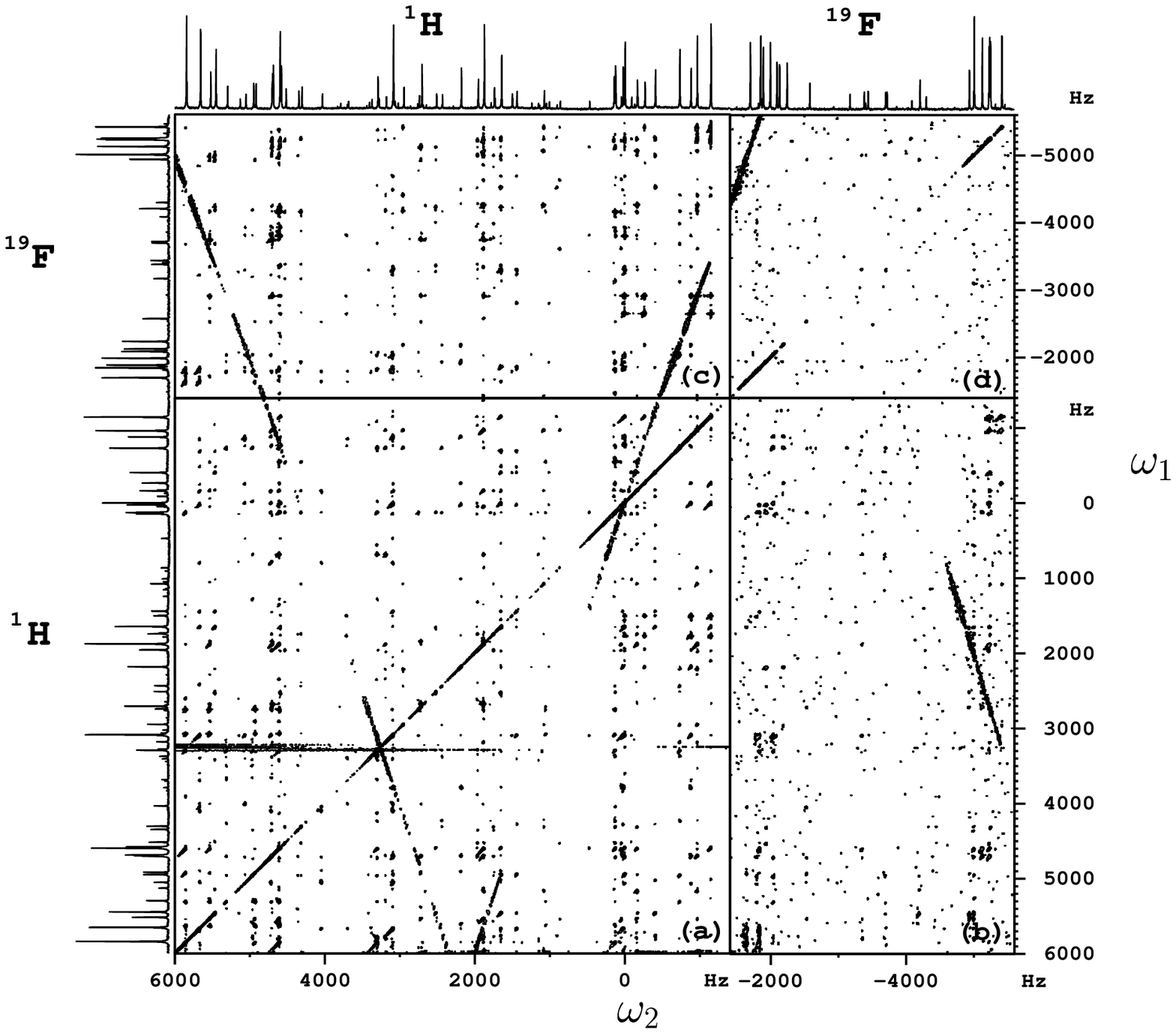}
\end{figure}
\hspace{5cm}
{\huge Figure 2}

\pagebreak
\begin{figure}
\epsfig{file=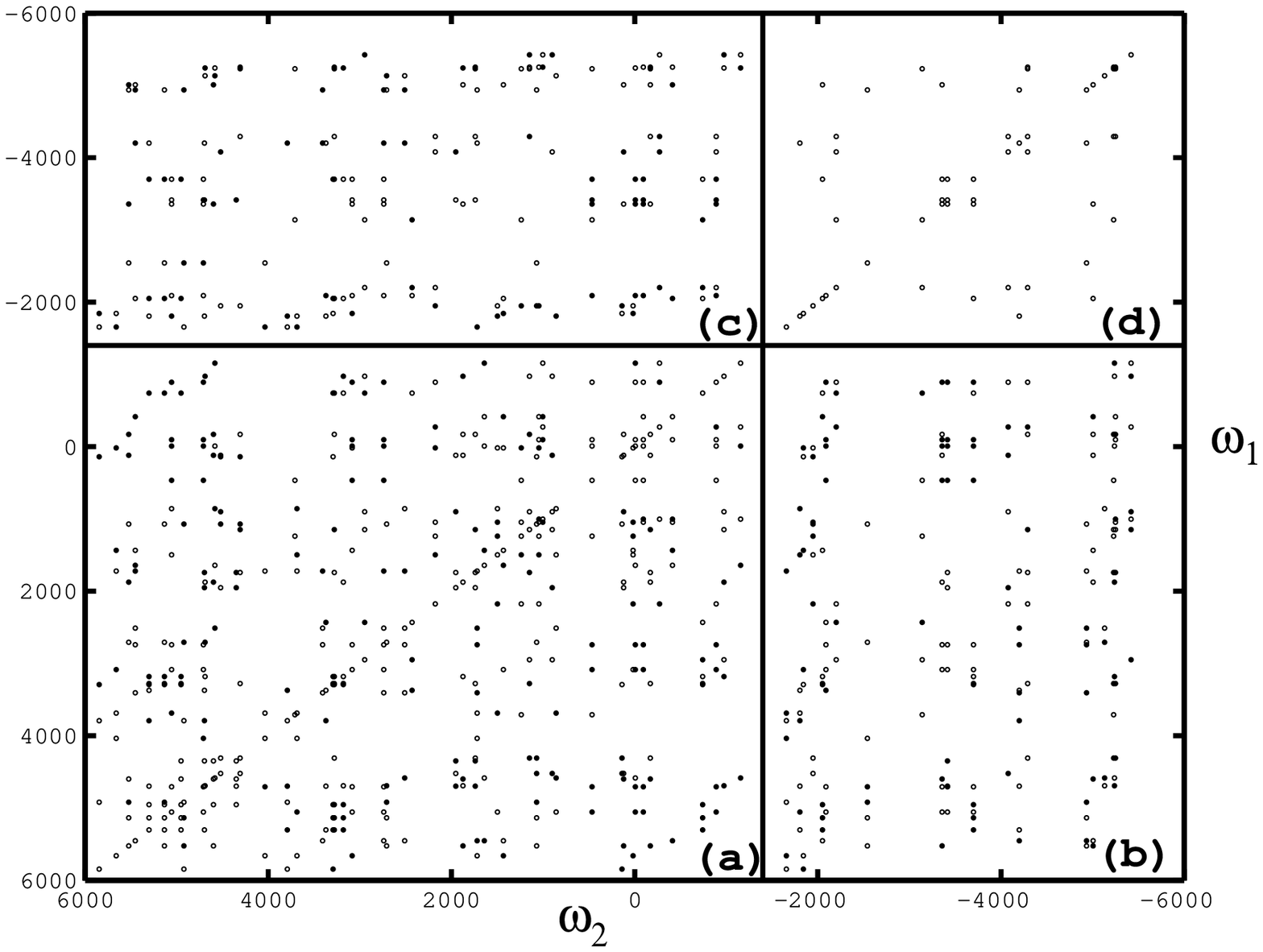,height=15cm,width=18cm}
\end{figure}
\hspace{5cm}
{\huge Figure 3}

\pagebreak
\vspace*{-3cm}
\begin{figure}
\epsfig{file=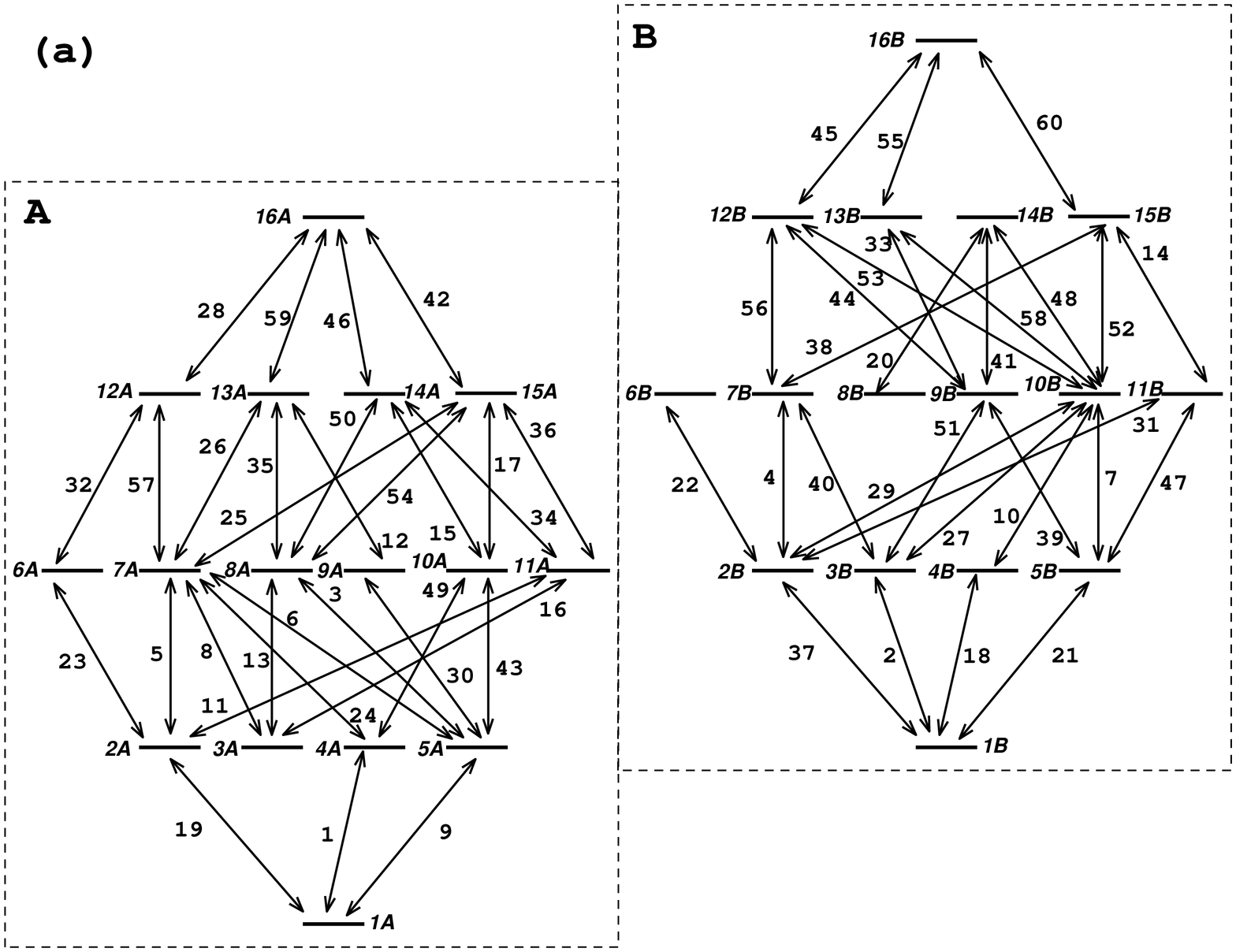,height=14cm,angle=270}
\end{figure}
\begin{figure}
\epsfig{file=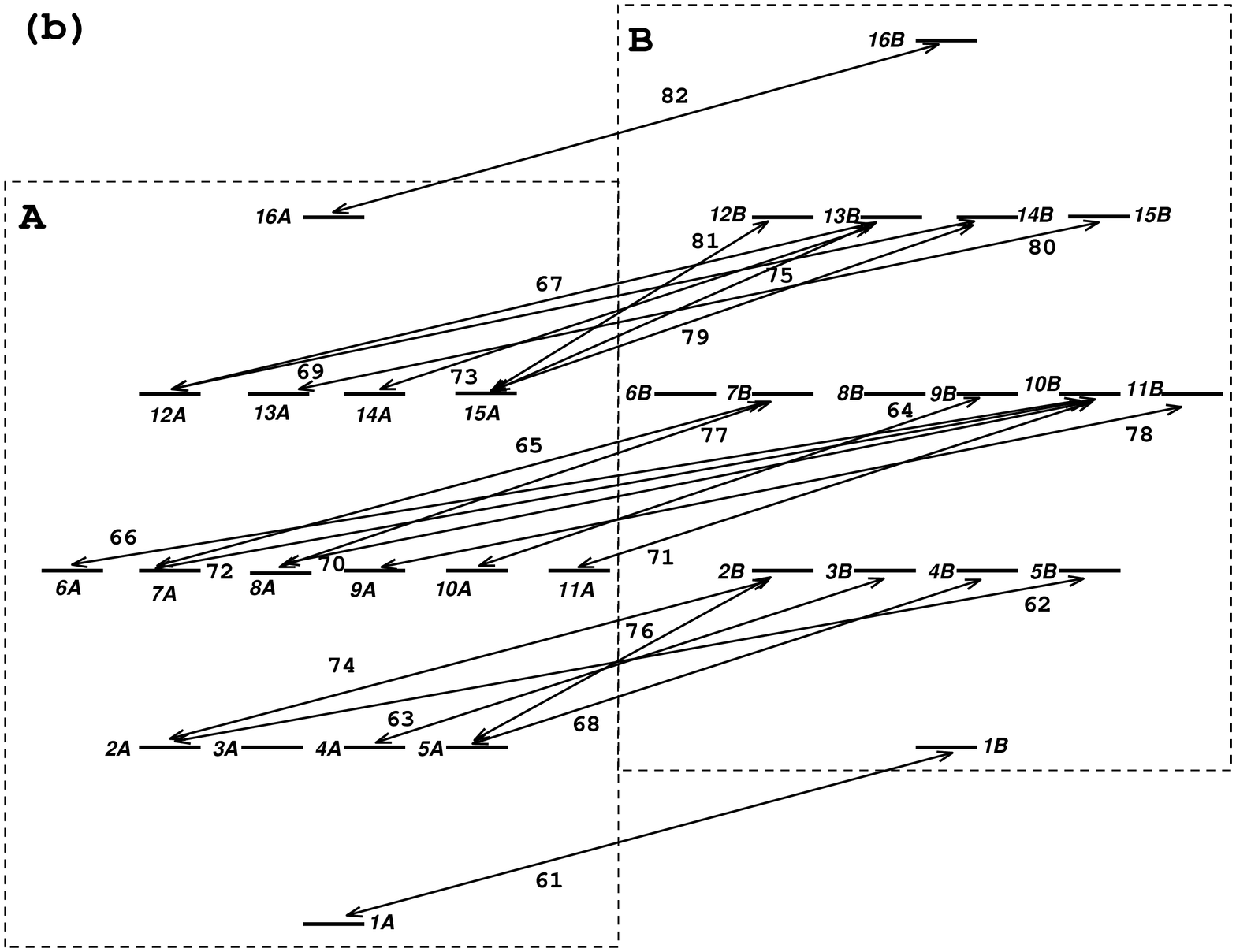,height=14cm,angle=270}
\end{figure}
\hspace{5cm}
{\huge Figure 4}

\pagebreak
\vspace*{-3cm}
\begin{figure}
\epsfig{file=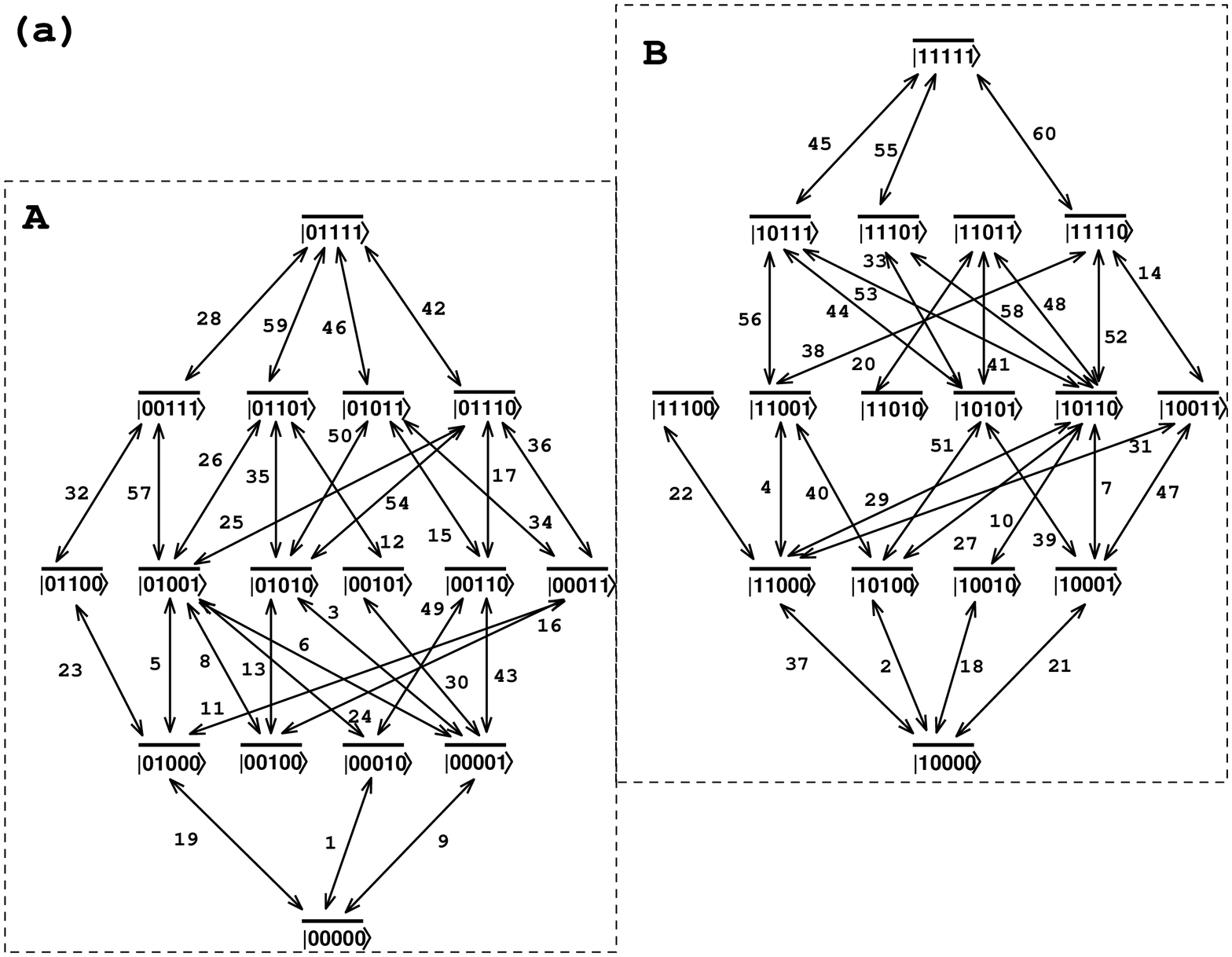,height=13cm,angle=270}
\end{figure}
\begin{figure}
\epsfig{file=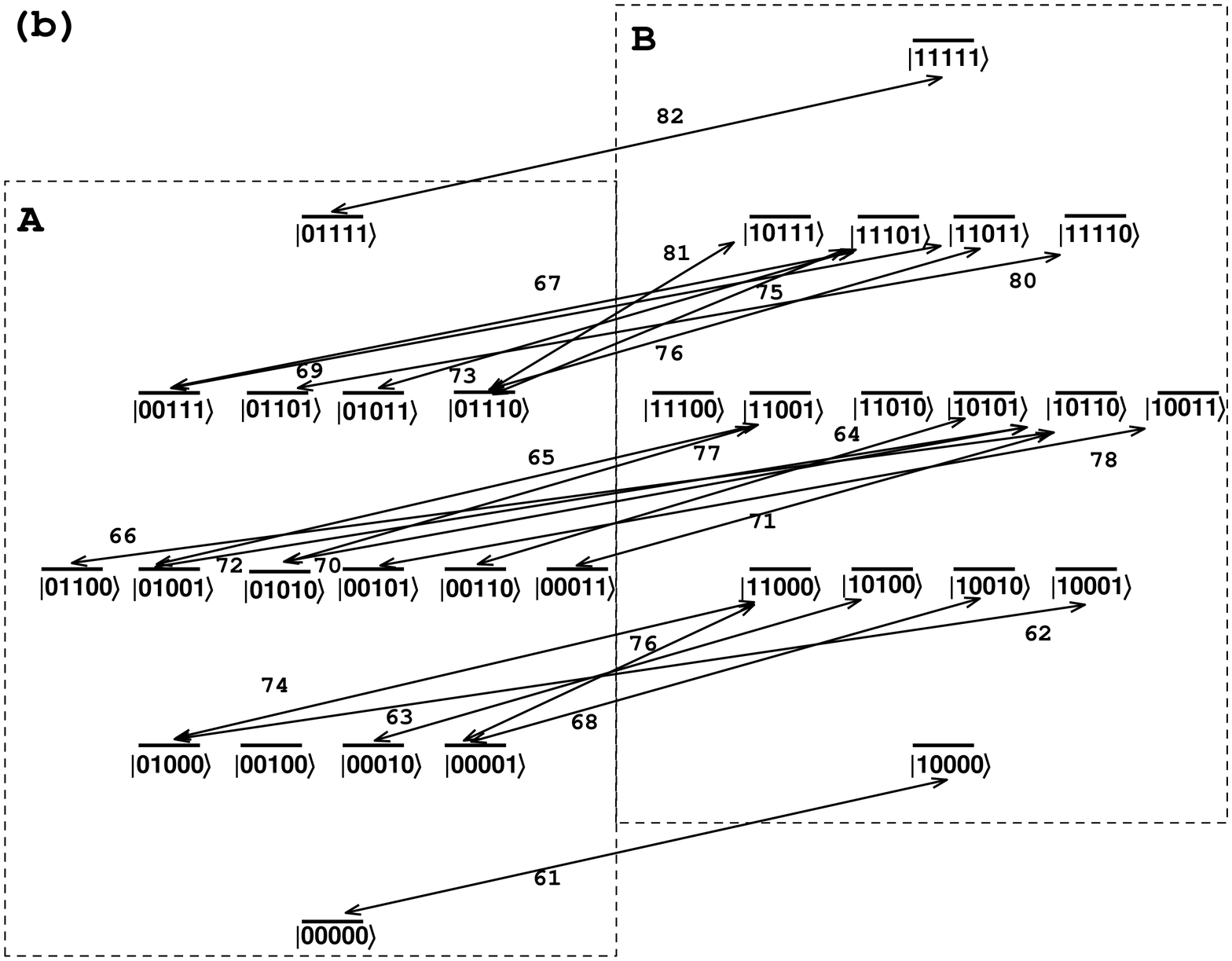,height=13cm,angle=270}
\end{figure}
\hspace{5cm}
{\huge Figure 5}

\pagebreak
\begin{figure}
\epsfig{file=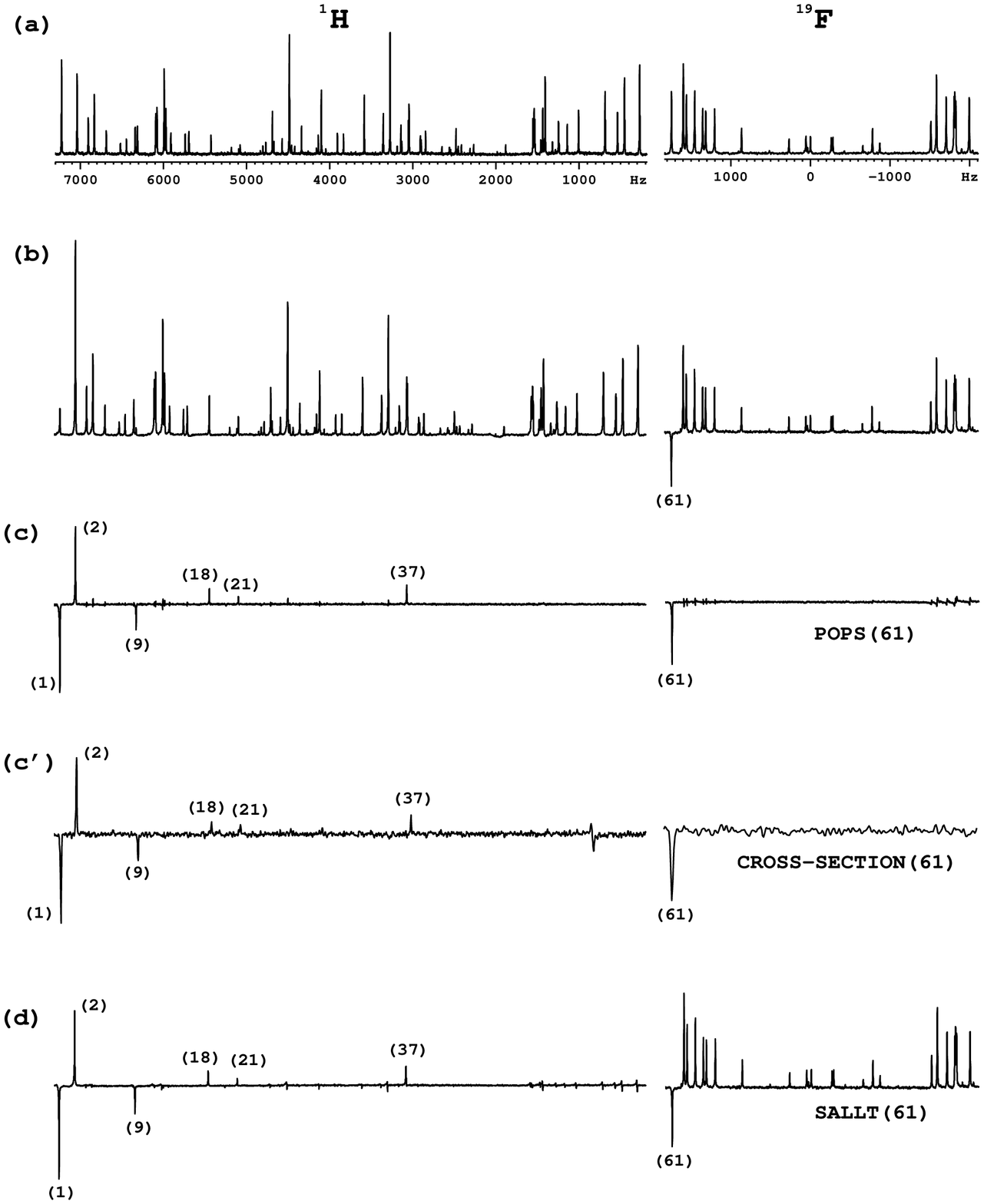,height=20cm}
\end{figure}
\hspace{5cm}
{\huge Figure 6}
\pagebreak
\begin{figure}
\epsfig{file=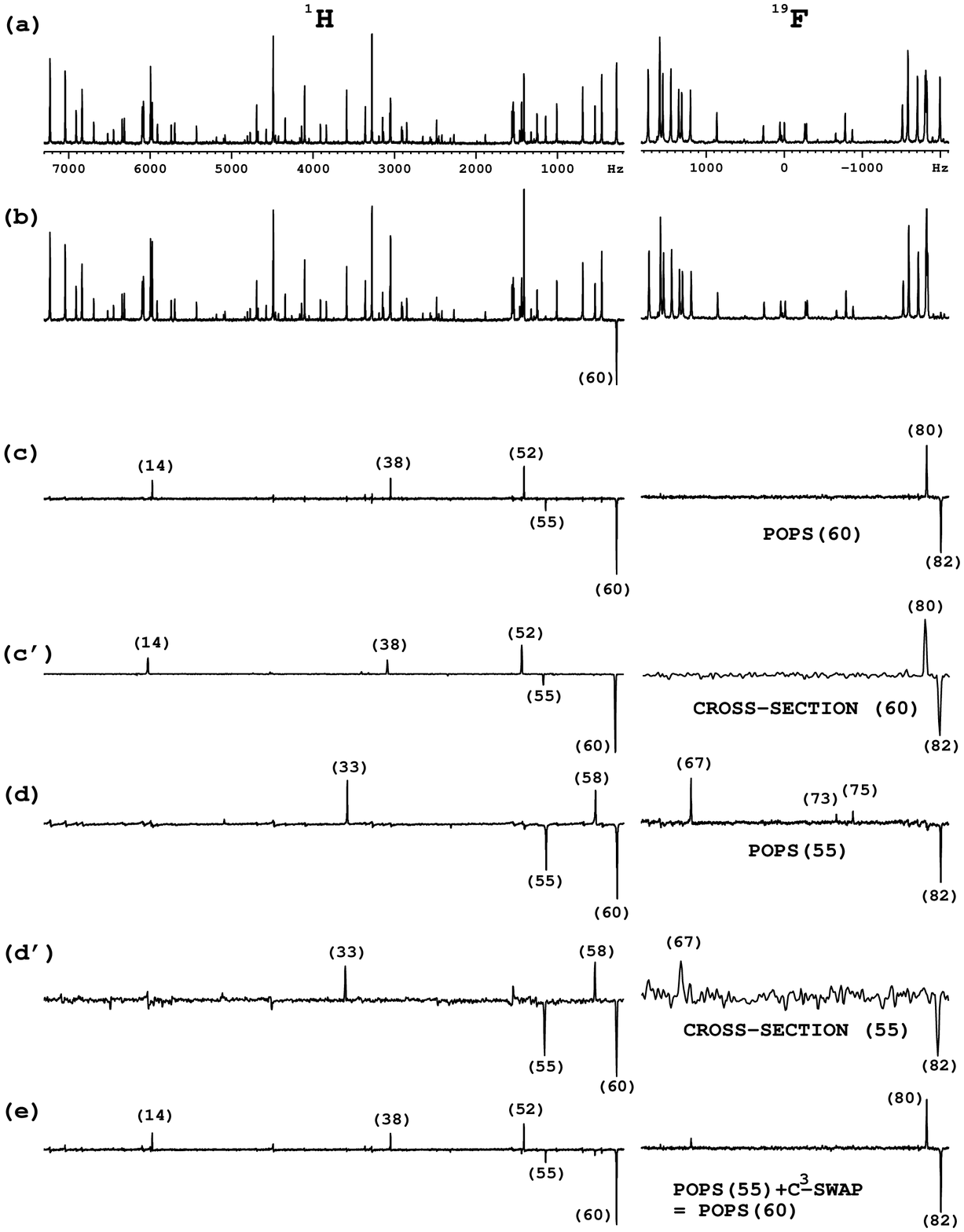,height=20cm}
\end{figure}
\hspace{5cm}
{\huge Figure 7}

\pagebreak
\vspace*{7cm}
\begin{figure}
\hspace{8cm}
\epsfig{file=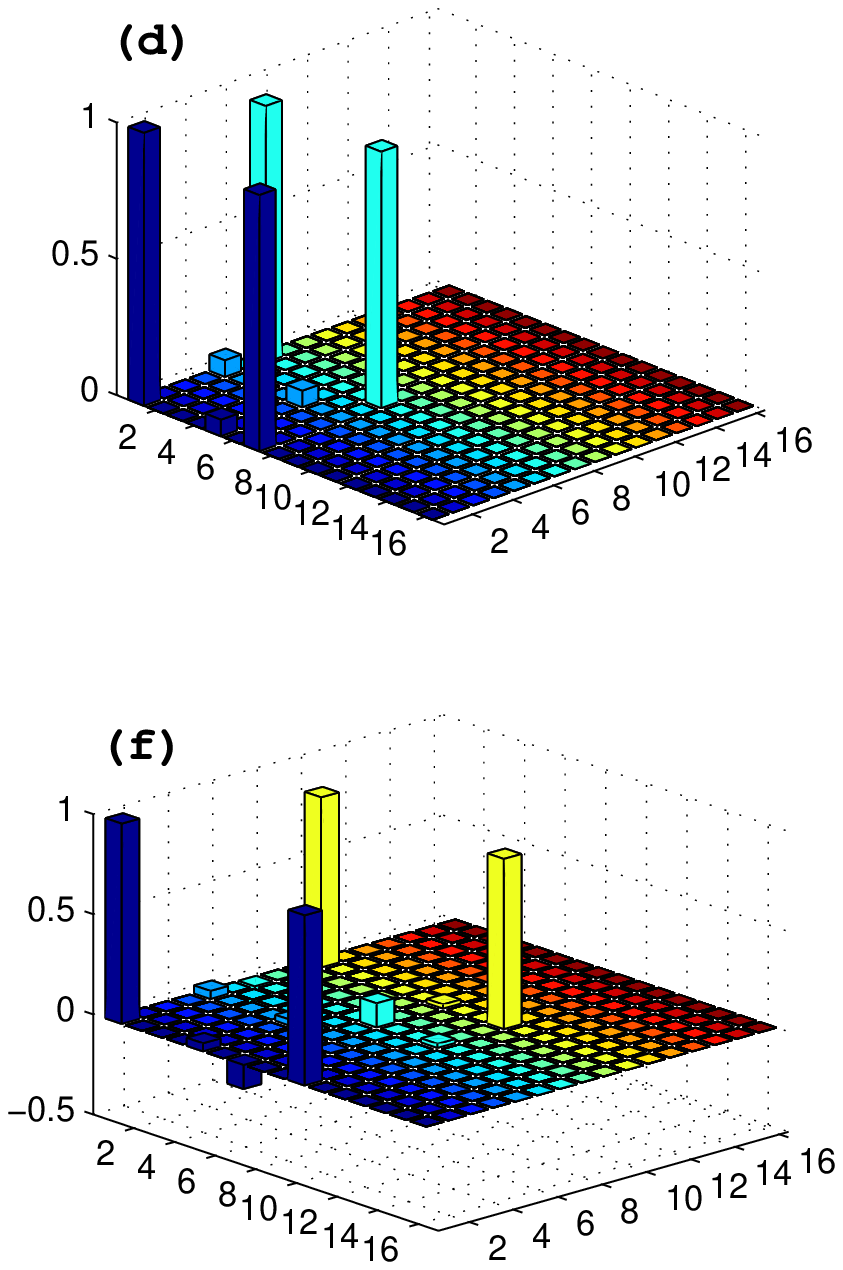,height=12cm,width=7cm}
\end{figure}
\vspace*{-20.5cm}
\begin{figure}
\epsfig{file=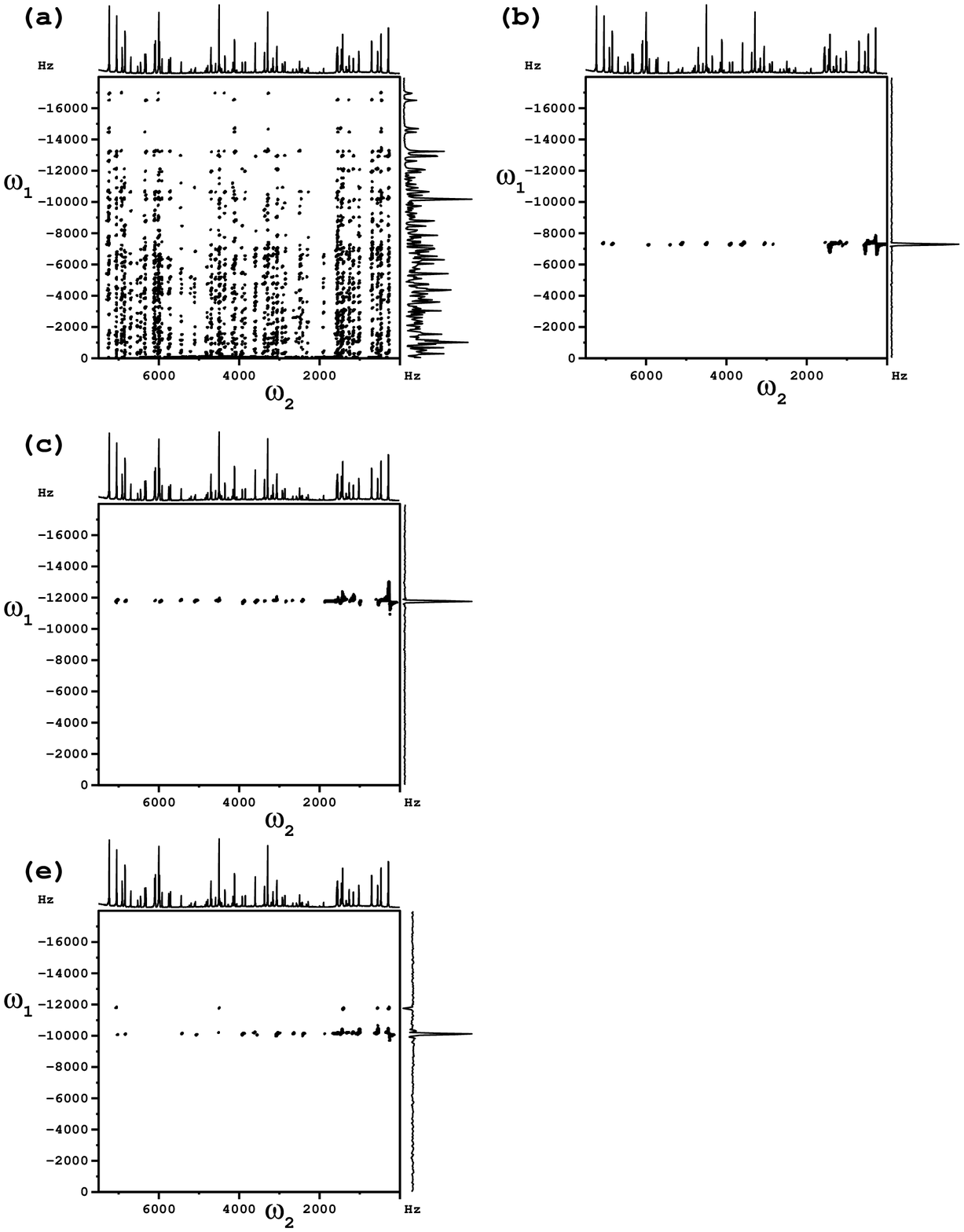,height=20cm}
\end{figure}
\hspace{5cm}
{\huge Figure 8}

\end{document}